\documentclass[aps,pre,reprint,superscriptaddress,doi=false,isbn=false,url=false,title=true,longbibliography,noeprint,footinbib]{revtex4-1}
\usepackage{graphicx}
\usepackage{amsmath}
\usepackage{array}
\usepackage{amsfonts}
\usepackage{float}
\usepackage{xcolor}
\usepackage{multirow}
\def\ba{\begin{eqnarray}}
\def\ea{\end{eqnarray}}
\def\be{\begin{equation}}
\def\ee{\end{equation}}
\def\bm{\begin{math}}
\def\me{\end{math}}

\newcommand{\dummy}

\makeatletter
\newcommand{\fmarki}{*}
\newcommand{\fmarkii}{\ensuremath{\dagger}}
\newcommand{\fmarkiii}{\ensuremath{\ddagger}}
\newcommand{\fmarkiv}{\ensuremath{\mathsection}}
\newcommand{\fmarkv}{\ensuremath{\mathparagraph}}
\newcommand{\fmarkvi}{\ensuremath{\|}}
\newcommand{\fmarkvii}{**}
\newcommand{\fmarkviii}{\ensuremath{\dagger\dagger}}
\newcommand{\fmarkix}{\ensuremath{\ddagger\ddagger}}
                
\def\@fnsymbol#1{{\ifcase#1\or \fmarki\or \fmarkii\or \fmarkiii\or \fmarkiv\or \fmarkv\or \fmarkvi\or \fmarkvii\or \fmarkviii\or \fmarkix \else\@ctrerr\fi}}
\makeatother

\begin{document}

\title{Temperature and Solvent Viscosity Tune the Intermediates During the Collapse of a Polymer}

\author{Suman Majumder}\email[]{smajumder.@amity.edu, suman.jdv@gmail.com}

\affiliation{Amity Institute of Applied Sciences, Amity University Uttar Pradesh, Noida 201313,
India
}
\author{Henrik Christiansen}\email[]{henrik.christiansen@itp.uni-leipzig.de}
\affiliation{Institut f\"{u}r Theoretische Physik, Universit\"{a}t Leipzig, IPF 231101, 04081 Leipzig, Germany 
}
\affiliation{NEC Laboratories Europe GmbH, Kurf\"{u}rsten-Anlage 36, 69115 Heidelberg, Germany}
\author{Wolfhard Janke}\email[]{wolfhard.janke@itp.uni-leipzig.de}
\affiliation{Institut f\"{u}r Theoretische Physik, Universit\"{a}t Leipzig, IPF 231101, 04081 Leipzig, Germany 
}
\date{\today}
\begin{abstract}
 Dynamics of a polymer chain in solution gets significantly affected by the temperature and the frictional forces arising due to solvent viscosity. Here, using an explicit solvent framework for polymer simulation with the liberty to tune the solvent viscosity, we study the nonequilibrium dynamics of a flexible homopolymer when it is suddenly quenched from an extended coil state in good solvent to poor solvent conditions. Results from our extensive simulations reveal that depending on the temperature $T$ and solvent viscosity, one encounters long-lived sausage-like intermediates following the usual pearl-necklace intermediates. Use of shape factors of polymers allows us to disentangle these two distinct stages of the overall collapse process, and the corresponding relaxation times. The relaxation time $\tau_s$ of the sausage stage, which is the rate-limiting stage of the overall collapse process, follows an anti-Arrhenius behavior in the high-$T$ limit, and the Arrhenius behavior in the low-$T$ limit. Furthermore, the variation of $\tau_s$ with the solvent viscosity provides evidence of internal friction of the polymer, that modulates the overall collapse significantly, analogous to what is observed for relaxation rates of proteins during their folding. This  suggests that the origin of internal friction in proteins is plausibly intrinsic to its polymeric backbone rather than other specifications.     
\end{abstract}

\maketitle

\section{INTRODUCTION}
Folding of a protein from its primary structure, a linear chain of a sequence of amino acids, to its three-dimensional native structure, is one of the most compelling research topics across all disciplines \cite{onuchic1997,shakhnovich2006,dill2008,dill2012,jumper2021,bhatia2022}. It is well known that the collapse of the polypeptide backbone of the protein molecule is an integral part of the overall folding process. Either it precedes or occurs simultaneously with the folding, eventually leading to the native conformation of the protein \cite{camacho1993,Pollack2001,Sadqi2003,haran2012,reddy2017}. 
Hence, over the years a lot of effort has been spent on investigating the collapse transition of general homopolymers and polypeptides in order to acquire useful insights into the overall folding process of proteins \cite{nishio1979,deGennes1985,yu1992,grosberg1993,byrne1995,timoshenko1995,kuznetsov1995,kuznetsov1996a,kuznetsov1996b,klushin1998,Halperin2000,hagen2000,abrams2002,montesi2004,Kikuch2005,xu2006,ye2007,pham2008,guo2011,udgaonkar2013,Majumder2015,Majumder2016,Majumder2017,christiansen2017,majumder2019,majumder2020,schneider2020}. 
\par
An extended polymer chain  undergoes a collapse or coil-globule transition when the solvent condition is changed from good (where the interaction between a monomer and solvent molecule is stronger) to poor (where the interaction among the monomers is stronger) \cite{rubenstein2003}. Before the advent of modern-day techniques like small-angle X-ray scattering, single-molecule fluorescence, dynamic light scattering, and dielectric spectroscopy, monitoring the collapse of a single polymer molecule experimentally was difficult. Consequently, in the past most studies concerning the dynamics of collapse relied on analytical theories and numerical simulations. In this regard, de Gennes was the first to propose a phenomenological model that describes formation of a sausage-like intermediate during the collapse, which eventually becomes a compact spherical globule via hydrodynamic dissipation of surface energy \cite{deGennes1985}. Following this, Dawson and co-workers performed a series of simulation studies \cite{byrne1995,timoshenko1995,kuznetsov1995,kuznetsov1996a,kuznetsov1996b} reporting a different sequence of events, which are in line with the phenomenological pearl-necklace picture of Halperin and Goldbart (HG) \cite{Halperin2000}. According to HG the collapse occurs via formation of an intermediate that resembles a pearl necklace where the pearls represent clusters of monomers formed along the chain. These clusters eventually grow bigger by coalescing with each other until a single cluster is left. The monomers within this single cluster further rearrange themselves to finally form a compact spherical globule. The observed phenomenology of a multi-stage collapse remains true irrespective of whether the simulations were performed with or without considering hydrodynamic effects \cite{Kikuch2005,Majumder2015,Majumder2017,schneider2020}.
\par
In all these previous studies, apart from the phenomenology of collapse, the focus was mainly on the  power-law scaling of the overall collapse time 
\begin{equation}
 \tau_c \sim N^{z_c},
\end{equation}
where $N$ is length of the linear polymer chain measured in terms of the number of monomers and $z_c$ is the corresponding dynamic exponent. Of course, the value of $z_c$ is dependent on the type of simulations performed. Simulations where hydrodynamic effects were not considered \cite{byrne1995,kuznetsov1995,kuznetsov1996a,kuznetsov1996b,Kikuch2005,pham2008,Majumder2017,christiansen2017} reported $1.5 \le z_c \le 2.0$. Consideration of hydrodynamic effects accelerates the dynamics, thus for those simulations \cite{kuznetsov1996a,Kikuch2005,pham2008,guo2011,schneider2020} the reported values are  $1.0 \le z_c \le 1.5$. The other aspect of the collapse is the scaling of the growth of clusters or pearls of monomers in the HG phenomenological picture. Even though this has been investigated in the past, only recently the growth of clusters of monomers has been understood using an analogy with the usual coarsening process popular in particle or spin systems \cite{Majumder2015,Majumder2017,christiansen2017}. In this regard, the associated aging and dynamical scaling has also been explored \cite{Majumder2016,majumder2016aging,Majumder2017,christiansen2017}.
\par
As discussed, the collapse transition of a polymer chain is attributed to the interaction between the solvent molecules and the monomers. Still, in most of the simulation studies mentioned above, the interaction of the polymer with the solvent molecules were considered implicitly. As a result, the model polymer exhibits a collapse transition driven by the temperature. It is known that such an approach provides static and dynamics properties of a polymer qualitatively compatible with both experimental and theoretical results. However, to disentangle the details of microscopic effects of solvent molecules on the nonequilibrium kinetics of the collapse transition, simulations with explicit solvent interaction is desired. Thus, there have been few studies using explicit solvent molecules both with or without hydrodynamics \cite{Kikuch2005,pham2008,reddy2017,guo2011,schneider2020}. Efforts include molecular dynamics (MD) simulations of the molecules, stochastic rotational dynamics \cite{malevanets1999}, and dissipative particle dynamics (DPD) \cite{hoogerbrugge1992,espanol1995,groot1997}. All of them again confirmed the same phenomenological pearl-necklace picture.
\par
Beside the solvent-polymer interaction, the collapse transition of a polymer gets influenced by other factors. The solvent viscosity is one such crucial factor. It may not only play a key role in the kinetics of the transition but can also alter the entire thermodynamic picture of collapse. As the polymer undergoes this conformational change, the solvent viscosity affects the diffusion of polymer segments, impacting the overall kinetics of the transition. Thus, understanding the effects of solvent viscosity on the collapse kinetics is of paramount importance for tailoring the design and performance of polymer-based materials. 
In this regard, for protein molecules it has been shown both in experiments and simulations that the kinetics of its overall folding gets significantly influenced while varying the viscosity of the solvent \cite{klimov1997,zagrovic2003,rhee2008,hagen2010,de2014,zheng2015}. Despite this obvious importance, to date no simulation studies have been performed to understand the effect of solvent viscosity on the overall kinetics of a simple homopolymer collapse. 
\par
Here, we present results from MD simulations of an explicit solvent polymer model where we employ the Lowe-Andersen (LA) thermostat to control the temperature \cite{lowe1999,koopman2006}. Our simulation set up not only enables the control of the interaction between the solvent molecules and the monomers of the polymer, but it also gives us the liberty to tune the viscosity of the solvent over a range that correspond to viscosities of real solvents like water. Our extensive simulations unravel the complex landscape of the pathways of collapse of a polymer in dependence of the solvent viscosity and temperature. It indicates that although in majority the  pearl-necklace scenario prevails, at low viscosity and temperature, one clearly observes the sausage-like intermediates proposed by de Gennes. By disentangling the scaling of the relaxation times associated with the different stages of the collapse along with the overall collapse time, we provide new insights regarding the temperature and solvent viscosity dependence of the kinetics.
\par
The remainder of the paper is organized in the following manner. Next, in Sec.\ \ref{methods}, we present details of the polymer model and an elaborate description of the employed simulation techniques. Following that, in Sec.\ \ref{results}, we present our results. Finally, in Sec.\ \ref{conclusion} we provide a brief summary of the main results and an outlook to future work.

\section{Model and Method}\label{methods}
\begin{figure}[t!]
\centering
\includegraphics*[width=0.5\textwidth]{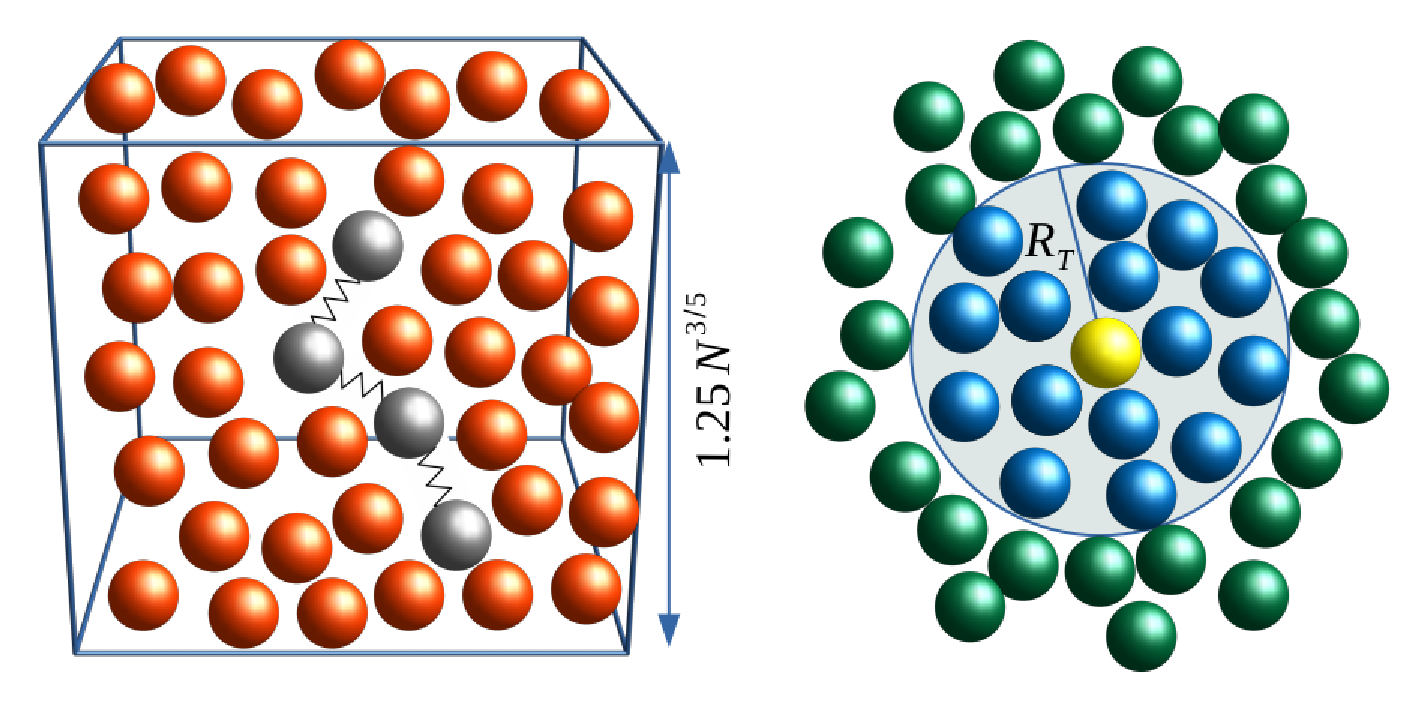}
\caption{\label{schematic} The schematic in the left panel shows a simple cubic box representing a typical simulation set up. The gray particles connected by springs represent the polymer chain and the orange particles represent the solvent molecules. The right panel shows how in the LA thermostat approach a particle chooses its partner for collision with the heat bath to achieve a new relative velocity. There the grey circle with radius $R_T$ around the central yellow particle marks the region within which any other particles (marked blue) will be considered for collision. Particles (marked green) that are beyond the distance $R_T$ will not be considered for collision with the chosen central particle.}
\end{figure}

We consider a bead-spring flexible homopolymer with explicit solvent molecules residing in a simple cubic box as shown in the schematic presented in the left panel of Fig.\ \ref{schematic}. The 
successive monomers at position $\vec{r}_i$ of the polymer chain are linearly connected  via a bond with energy given by the standard 
finitely extensible non-linear elastic (FENE) potential 
\begin{eqnarray}\label{FENE}
E_{\rm{FENE}}(r_{i,i+1})=-\frac{K}{2} R^2 \ln \left[ 1-\left( \frac{r_{i,i+1}-r_0}{R} \right)^2 \right],
\end{eqnarray}
where $r_{i,i+1}=|\vec{r}_{i+1}-\vec{r}_i|$, the spring constant $K=40$, equilibrium bond length $r_0=0.7$, and its maximum extension $R=0.3$. Both the monomer and solvent molecules are assumed to be spherical 
beads with diameter $\sigma$ and mass $m\equiv1$. All nonbonded interactions, i.e., solvent-solvent (s-s), solvent-monomer (s-m), 
and monomer-monomer (m-m) interactions are given  by 
\begin{equation}\label{LJ_rc}
 E_{\text{nb}}(r_{i,j})=E_{\text{LJ}}(r_{i,j})-E_{\text{LJ}}(r_c)-(r_{i,j}-r_c)\left(\frac{\partial E_{\text{LJ}}}{\partial r_{i,j}}\right)_{r_{i,j}=r_c},
\end{equation}
where $E_{\text{LJ}}(r)$ is the standard Lennard-Jones (LJ) potential given as 
\begin{equation}\label{LJ_pot}
 E_{\text{LJ}}(r)=4\epsilon \left [ \left(\frac{\sigma}{r} \right)^{12} - \left( \frac{\sigma}{r} \right)^{6} \right].
\end{equation}
We set $\sigma =r_0/2^{1/6}$ and  the interaction strength $\epsilon\equiv1$. To facilitate the collapse transition of the polymer we choose the cutoff radius $r_c$ in Eq.\ \eqref{LJ_rc} in the following way 
\begin{equation}\label{rc_choice}
 r_c=
\begin{cases}
  2^{1/6}\sigma~~~ \text{for s-s and s-m interaction }\\
  2.5\sigma~~~~~ \text{for m-m interaction}\,
   \end{cases}.
\end{equation}
This implies a purely repulsive potential for s-s and m-m interactions, and a short-range attractive potential for the m-m interaction, mimicking a bad solvent condition for a polymer. On the other hand, a choice of purely repulsive potential by setting $r_c=2^{1/6}\sigma$ for all the above three  types of interactions mimics good solvent conditions.
\par
We carry out MD simulations at a constant temperature using the LA 
thermostat. There, the position $\vec{r}_i$ and velocity $\vec{v}_i$ 
of the $i$-th bead is updated using Newton's equations  
\begin{equation}\label{newton1}
 \frac{d\vec{r}_i}{dt}=\vec{v}_i;~~
  \frac{d\vec{v}_i}{dt}=\vec{f}_i,
\end{equation}
where $\vec{f}_i$ is the conservative force (originating from the bonded and nonbonded interactions) acting 
on the particle. The equations of motion are solved using the velocity-Verlet integration scheme. Up to this point the method represents the usual microcanonical MD \cite{Frenkel_book}. To control the temperature using  
the LA thermostat \cite{lowe1999,koopman2006}, one considers a pair of particles within a distance $R_T$ as shown in the schematic presented in the right panel of Fig.\ \ref{schematic}. A heat-bath collision of this pair of particles with a probability $\Gamma \Delta t$ is then performed  by assigning a new relative velocity to the pair, drawn from the Maxwellian distribution. Here, $\Delta t$ is the time step and $\Gamma$ is 
the collision frequency. The update of relative velocities is only done on the component parallel to the line joining the centers of the pair of particles, thus conserving the angular momentum. Additionally, the new velocities are distributed to the chosen pair in such a way that the linear momentum is also conserved. This makes the thermostat Galilean invariant and hence preserves hydrodynamic effects \cite{lowe1999,koopman2006}.

\par
The LA approach is an alternative approach to DPD, however, with the liberty to use large $\Delta t$ in the velocity-Verlet integration scheme \cite{marsh1997,nikunen2003}. The other 
advantage is that by varying $R_T$ and $\Gamma$ one can tune the bath collision frequency, i.e., effectively 
controlling the frictional drag or, in other words, the solvent viscosity. We fix $R_T= 2^{1/6}\sigma$ and vary $\Gamma$ over a wide range with the goal to cover solvents of diverse 
viscosity. This choice of $R_T$ reproduces the known equilibrium properties of a polymer \cite{majumder2019lowe}.
\par
Following the usual protocol, first, we generate a random walk of length $N$ on a simple-cubic lattice and then put this walk or chain in a 
box of length $L=1.25N^{3/5}$. Subsequently, we insert solvent particles at random positions keeping the density fixed to $\rho=0.7$ and ensure that the solvent particles do not overlap 
with the monomers of the polymer chain. Then we run our simulations at the desired temperature in good solvent condition [with $r_c=2^{1/6}\sigma$ in Eq.\ \eqref{LJ_rc}] for $10^7$ MD steps using $\Delta t=0.005 \tau_0$ 
to get an equilibrated polymer with extended conformation. In our simulations, the unit of temperature $T$ is $\epsilon/k_B$ (where we chose $k_B=1$) and the unit of time 
is the standard LJ time unit $\tau_0=(m\sigma^2/\epsilon)^{1/2}$. Since we are interested in the collapse kinetics of the polymer, once the system is equilibrated, we change the solvent condition to poor by fixing $r_c$ according to Eq.\ \eqref{rc_choice}, and let the system evolve as we simultaneously measure various physical quantities that will be presented subsequently. All other parameters remain the same as already stated.

\par
It is easy to perceive that by varying $\Gamma$ in our simulations one can tune the frictional drag or effectively the solvent viscosity. Hence, subsequently in the rest of the paper we use $\Gamma$ to  represent the solvent viscosity. We systematically vary $\Gamma$  at different $T$ with the goal to encompass a wide range of solvent viscosity.  For faster computation we use LAMMPS \cite{plimpton1995} which we have already modified to implement the LA thermostat \cite{majumder2019lowe}. We choose polymers of chain length $N \in [128,2048]$ for which the number of solvent particles is roughly the in the range $[8.5\times 10^3,1.25\times 10^6]$. Except for the snapshots all results presented are averaged over at least $50$ different independent initial and thermal realizations generated using different seeds of the random number generator. 
\begin{figure*}[t!]
\centering
\includegraphics*[width=0.95\textwidth]{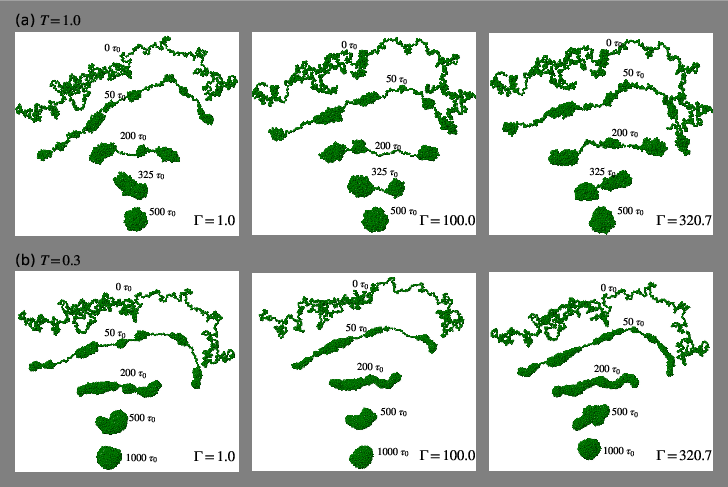}
\caption{\label{snapshots} Snapshots at different times representing the evolution of a flexible homopolymer of length $N=1024$, following a quench from an extended state in a good solvent to a globular state in poor solvent condition. Results for two different temperatures $T$ are presented as indicated. At each temperature results for three different solvent viscosities $\Gamma$ are shown.}
\end{figure*}
\begin{figure*}[t!]
\centering
\includegraphics*[width=0.95\textwidth]{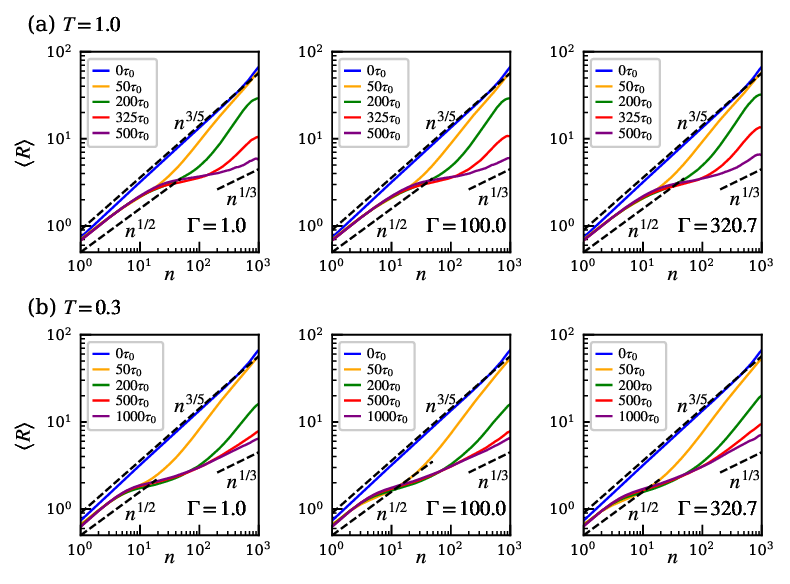}
\caption{\label{cont_dist} Double-log plots of the mean Euclidean distance $\langle R \rangle$ among two monomers as a function of the contour distance $n$ between them along the chain, at two different temperatures $T$ for different solvent viscosities $\Gamma$, as indicated. For each case data at times corresponding to the snapshots in Fig.\ \ref{snapshots} are presented. The dashed black lines representing $\sim n^{3/5}$, $n^{1/2}$, and $\sim n^{1/3}$, respectively, correspond to the expected equilibrium behaviors for a self-avoiding, Gaussian  and collapsed polymer.}
\end{figure*}
\section{Results}\label{results} 
We present the main results by subdividing them into four subsections. In the first subsection, we discuss the overall phenomenology of the collapse, and corresponding analyses required to extract various relaxation times. In the next one we present the scaling of various relaxation times with respect to chain length. In the following two subsections we present, respectively, the temperature and viscosity dependence of various relaxation times.
\subsection{Collapse phenomenology and relaxation times}\label{phenomenology}
The phenomenology of the collapse following a solvent quench can be understood from the time evolution snapshots, illustrated in Fig.\ \ref{snapshots} for a polymer of length $N=1024$ at two judicially chosen values of the temperature $T$ \footnote{Traditionally, the choice of $T=1.0$ is a more natural choice. Hence in all the figures and tables the data for $T=1.0$ are always presented first. Also, the otherwise unusual choice of $\Gamma=320.7$ is due to the fact that with our choice of integration time step $\Delta t=0.005 \tau_0$, for $\Gamma=320.7$ the collision probability of the particles is maximum, i.e., $1$.}. For each $T$ we show the time evolution for three different values of the solvent viscosity $\Gamma$, as indicated. In each case we present typical snapshots at different times starting from an extended coil at $t=0$ to a time where a steady globule is observed. In each case the collapse initiates with the formation of local clusters of monomers along the chain, as shown by the snapshots at $t=50\tau_0$. Subsequently these local clusters merge with each other to form bigger clusters and this continues until a single cluster is left. This apparently agrees with the phenomenological picture proposed by HG \cite{Halperin2000}. However, a careful inspection of the snapshots across all the presented values of $T$ and $\Gamma$ reveals that only for the higher $T$ case, irrespective of $\Gamma$, the local clusters of monomers represent true spherical clusters as expected in HG's pearl-necklace picture. During the course of the collapse at one stage inevitably the polymer attains a dumbbell
conformation when only two clusters are present, as represented in Fig.\ \ref{snapshots}(a) by the snapshot at $t=325 \tau_0$  for $\Gamma=100.0$. Eventually, these two clusters also coalesce with each other and the globule attains an approximate spherical shape.  
\par
For the low-$T$ case, on the other hand, we observe that the local clusters formed at the initial stages of the collapse do not represent a spherical pearl but rather a sausage-like shape as $\Gamma$ increases. Thus, the chain resembles a sausage necklace rather than a pearl necklace. Subsequently, the connected sausages coalesce with each other to form longer sausages, and eventually when the last two sausages coalesce the polymer assumes a single cylindrical sausage-like conformation as suggested by de Gennes \cite{deGennes1985}. Following this, the polymer eventually deforms in a slow rearrangement process from the sausage-like conformation to a spherical globule shape. Overall, the inspection of the time evolution snapshots implies that the phenomenology of collapse is dependent on temperature $T$ as well as on the solvent viscosity $\Gamma$. At high $T$, irrespective of solvent viscosity the phenomenology is purely described by the pearl-necklace picture of HG. On the other hand, at low $T$ it is dependent on the solvent viscosity $\Gamma$ and one observes sausage-like clusters as an intermediate before the polymer attains a spherical shape. 

\par
For the first quantitative understanding of the time evolution of the polymer during the collapse we calculate the Euclidean distance between two monomers $i$ and $j$ as
\begin{equation}
 R=|\vec{r}_i-\vec{r}_j|.
\end{equation}
 The variation of $R$ is then noted  as a function of the contour distance 
\begin{equation}
 n=|i-j|.
\end{equation}
 Such plots of $R$ at times corresponding to the snapshots in Fig.\ \ref{snapshots} are presented in Fig.\ \ref{cont_dist}. There $\langle \dots \rangle$ denotes averaging over different simulation runs and all possible pairs of monomers. It is expected that $\langle R \rangle$ obeys a power-law scaling of the form \cite{deGennesbook,rubenstein2003}
 \begin{equation}\label{R_scaling}
  \langle R \rangle \sim n^{\nu}.
 \end{equation}
The data for $t=0$, i.e., the starting configuration in all cases of Fig.\ \ref{cont_dist} are consistent with Eq.\ \eqref{R_scaling} with an exponent $\nu=3/5$, expected for an extended polymer chain in good solvent \cite{deGennesbook}. Once the solvent condition is changed to poor, the data at large $n$ gradually start deviating from the $\langle R \rangle \sim n^{3/5}$ behavior represented by the upper dashed line. At the latest time for large $n$ in all cases the data seem to be quite consistent with Eq.\ \eqref{R_scaling} with $\nu=1/3$, expected for a globular conformation  \cite{deGennesbook}. At large $t$, for small $n$ the data seem to be consistent with $\langle R \rangle \sim n^{1/2}$, i.e., a Gaussian behavior. This can be understood by considering the pearls or globules to be made up of several small segments of the polymer itself. Each of these segments feels the highly dense surrounding consisting of all the other segments, effectively mimicking a polymer-melt like environment where a Gaussian behavior is expected \cite{deGennesbook}. The agreement of the data with the expected scaling at early and late times confirm the formation of globule-like conformation through collapse of an extended coil-like conformation, irrespective of the temperature $T$ and solvent viscosity $\Gamma$. However, no signature of the different collapse pathways as identified from the time evolution snapshots in Fig.\ \ref{snapshots} is resolved.
\begin{figure}[t!]
\centering
\includegraphics*[width=0.45\textwidth]{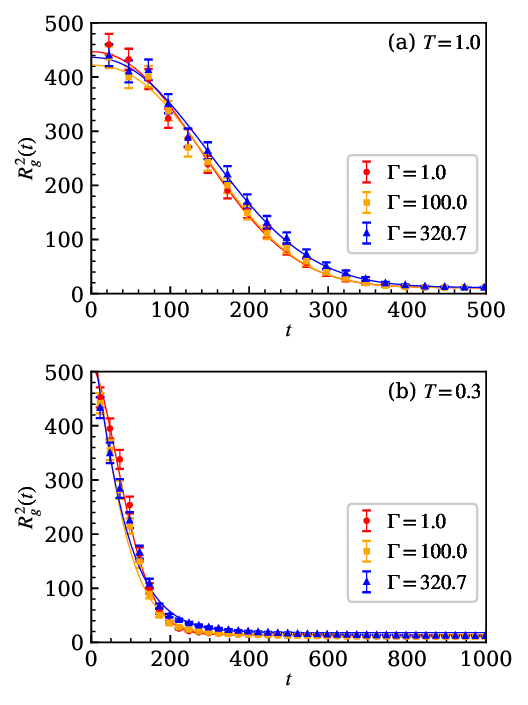}
\caption{\label{Rg_plot}  Time dependence of the squared radius of gyration $R_g^2$ during the collapse of a polymer of length $N=1024$ at (a) $T = 1.0$ and (b) $T=0.3$ for different $\Gamma$ as indicated. The solid lines represent the best fit curves obtained by fitting the data with the stretched exponential decay prescribed in Eq.\ \eqref{Rg_fit}. The results obtained from the fitting exercise are quoted in Table\ \ref{fit_Rg}.}
\end{figure}
\begin{table}[b!]
\caption{Fitting results for the time dependence of $R_g^2$ data at different $T$ and $\Gamma$ using the ansatz in Eq.\ \eqref{Rg_fit}.}
\begin{center}
\begin{tabular}{m{1.0cm} m{1.0cm} m{1.0cm} m{1.25cm} m{1.25cm} m{1.0cm}}
    \hline
    \hline
    \vskip 0.1cm
    $T$&$\Gamma$ &$b_0$ &~~~~~$b_1$ &~~~~$\tau_{\rm f}$ &~~~~$\beta$ \\
    \hline
    \hline
    \vskip 0.1cm
    \multirow{3}{*}{\rotatebox[origin=c]{0}{$1.0$}}~~~~~~~~~~~~$1.0$ & ~~~~~~~~~~~~~$10.81(4)$ &~~~~~~~~~~~~~~~~~~~~~$436(9)$ &~~~~~~~~~~~~~~~~~~~~~~~$184(3)$ &~~~~~~~~~~~~~~~~~~~~~~$2.18(5)$ \\  
  ~~~~~~~~~~~~~~$100.0$ & ~~~~~~~~~~~~~$10.90(3)$ &~~~~~~~~~~~~~~~~~~~~~$411(9)$ &~~~~~~~~~~~~~~~~~~~~~~~$191(3)$ &~~~~~~~~~~~~~~~~~~~~~~$2.32(6)$ \\ 
  ~~~~~~~~~~~~~~$320.7$ & ~~~~~~~~~~~~~$11.33(9)$ &~~~~~~~~~~~~~~~~~~~~~$425(7)$ &~~~~~~~~~~~~~~~~~~~~~~~$199(3)$ &~~~~~~~~~~~~~~~~~~~~~~$2.17(5)$ \\
   \hline
   \vskip 0.1cm
    \multirow{3}{*}{$0.3$}~~~~~~~~~~~~$1.0$ & ~~~~~~~~~~~~~$13.98(33)$ &~~~~~~~~~~~~~~~~~~~~~$496(41)$ &~~~~~~~~~~~~~~~~~~~~~~~$106(8)$ &~~~~~~~~~~~~~~~~~~~~~~$1.70(9)$ \\  
  ~~~~~~~~~~~~~~$100.0$ & ~~~~~~~~~~~~~$13.43(34)$ &~~~~~~~~~~~~~~~~~~~~~$565(68)$ &~~~~~~~~~~~~~~~~~~~~~~~~~$80(10)$ &~~~~~~~~~~~~~~~~~~~~~~$1.22(9)$ \\ 
  ~~~~~~~~~~~~~~$320.7$ & ~~~~~~~~~~~~~$18.09(79)$ &~~~~~~~~~~~~~~~~~~~~~$554(67)$ &~~~~~~~~~~~~~~~~~~~~~~~~~$84(10)$ &~~~~~~~~~~~~~~~~~~~~~~$1.13(8)$ \\
    \hline
\end{tabular}
\end{center}
\label{fit_Rg}
\end{table}
\par
Following the traditional approach to monitor the kinetics we also calculate the squared radius of gyration
\begin{equation}\label{Rg}
 R_g^2=\frac{1}{2N^2}\sum_{i,j}(\vec{r}_i-\vec{r}_j)^2.
\end{equation}
The corresponding time dependence of $R_g^2$ is presented in Fig.\ \ref{Rg_plot} for the systems presented in Figs.\ \ref{snapshots} and \ref{cont_dist}. At both $T$, the data for different $\Gamma$ apparently do not imply any general trend. Also, the decay of $R_g^2$ with time does not provide clear signature of the intermediate stages of the collapse.  Nevertheless, we extract a relaxation time $\tau_{\rm f}$ proportional to the overall collapse time by fitting the curves with the following stretched exponential function 
 \begin{equation}\label{Rg_fit}
  R_g^2(t)=b_0+b_1\exp\left[ -\left(\frac{t}{\tau_{\rm f}}\right)^{\beta}\right],
\end{equation}
where $b_0$ corresponds to $R_g^2$ of the collapsed conformation, and $b_1$ and $\beta$ are associated nontrivial fitting parameters. The results of the fitting exercises on the data presented in Fig.\ \ref{Rg_plot} are tabulated in Table\ \ref{fit_Rg}. The obtained relaxation times $\tau_{\rm f}$ for both $T$ have no clear systematic trend as a function of $\Gamma$. The value of $\tau_{\rm f}$ indicates a faster collapse at $T=0.3$, contradicting with the actual fact  revealed from the time evolution snapshots in Fig.\ \ref{snapshots} which gives a clear indication that the overall collapse time, i.e., the time it takes to form a spherical globule from an extended coil, increases as $T$ decreases. Also note the difference in the values of $b_0$, which correspond to the respective equilibrium value of $R_g^2$ at the two $T$. At $T=0.3$ one notices a $20\%$ drop in the value of $\tau_{\rm f}$ as solvent viscosity changes from $\Gamma=1.0$ to $\Gamma=100.0$, suggesting a faster collapse in viscous solvent. This is also not in agreement with the picture reflected by the time evolution snapshots at $T=0.3$ in Fig.\ \ref{snapshots}. Furthermore, the values of $\beta \in [1.13,2.32]$ quoted in Table\ \ref{fit_Rg} are also significantly higher than the previously reported values obtained from Monte Carlo simulations of a homopolymer collapse without explicit solvent  \cite{Majumder2017,christiansen2017}. Overall, it can be inferred that the traditional approach of extracting the relaxation times using the decay of $R_g^2$ is not unambiguous in the present case. Importantly, it cannot be used to draw any information on the time scales that separate different stages of the collapse, especially the transition from the pearl-necklace-like to the sausage-like picture at low $T$.  

\par
The variety of conformations observed for a polymer during its collapse as presented in Fig.\ \ref{snapshots} motivates us now to investigate various shape properties of the polymer. For equilibrium conformations of polymer, shape properties have proven to be  useful tools in identifying various structural transitions \cite{alim2007,ostermeir2010,blavatska2010,arkin2013}. With this motivation we first calculate the full gyration tensor ${\bf Q}$ having the components
\begin{equation}
 Q_{ij}=\frac{1}{N}\sum_{k=1}^N(x_k^i-x_{\rm CM}^i)(x_k^j-x_{\rm CM}^j),~i,j=1,2,3
\end{equation}
where $x_{\rm CM}^i$ is the $i$-th component of the center-of-mass vector
\begin{equation}
 \vec{r}_{\rm CM}=\frac{1}{N}\sum_{i=1}^N \vec{r}_i.
\end{equation}
% Note that the already calculated squared radius of gyration $R_g^2$ can also be obtained from ${\bf Q}$ as 
% \begin{equation}
%  R_g^2={\rm Tr}\ {\bf Q}=\sum_{i=1}^3Q_{ii}.
% \end{equation}
% Since, we have already checked that $R_g^2$ is insufficient to disentangle various stages of the collapse 
We move onto calculating various shape properties derived from ${\bf Q}$. The spread in eigenvalues $\lambda_i$ of ${\bf Q}$ gives an idea about how uniformly the monomers of the polymer are distributed spatially, and thus measures the asymmetry of the chain. Various combination of the eigenvalues provide different useful measures of the asymmetry present in the polymer conformation. To proceed further, we first calculate the mean eigenvalue 
\begin{equation}
 \bar{\lambda}= \frac{{\rm Tr}\ {\bf Q}}{3}=\frac{1}{3}\sum_{i=1}^3 \lambda_i= \frac{R_g^2}{3}.
\end{equation}
This allows us to calculate the extent of asphericity of a conformation \cite{theodorou1985,aronovitz1986,cannon1991,blavatska2010,arkin2013}
\begin{equation}
 A_3=\frac{3}{2}\frac{{\rm Tr}\ {\bf \tilde{Q}^2}}{({\rm Tr}\ {\bf \tilde{Q}})^2}=\frac{1}{6}\sum_{i=1}^3\frac{(\lambda_i-\bar{\lambda})^2}{\bar{\lambda}^2},
\end{equation}
where ${\bf \tilde{Q}}={\bf Q}-\bar{\lambda}{\bf I}$ with ${\bf I}$ being the unity matrix. $A_3$ can be used to distinguish shapes of different polymer conformations. For example $A_3=0$ for a completely spherical conformation and $A_3=1$ for perfectly rod-like conformation. For any polymer conformation it obeys the bounds $0\le A_3 \le1$. We also calculate the prolateness of a polymer conformation as \cite{aronovitz1986,blavatska2010,arkin2013}
\begin{equation}
 S= 27 \frac{{\rm det {\bf \tilde{Q}}}}{({\rm Tr}\ {\bf \tilde{Q}})^3}=\frac{\prod_{i=1}^3(\lambda_i-\bar{\lambda})}{{\bar{\lambda}}^{3}}.
\end{equation}
For an absolute prolate conformation, i.e., for a rod $\lambda_1\ne0$, $\lambda_2=\lambda_3=0$ implying that $S=2$. For an absolute oblate conformation, i.e., for a disk $\lambda_1=\lambda_2$, $\lambda_3=0$, implying $S=-1/4$. Thus, the bounds are $-1/4\le S\le 2$. We also calculate another quantity defined as the nature of asphericity \cite{cannon1991,alim2007,ostermeir2010}
\begin{equation}
 \Sigma=\frac{4 {\rm det} {\bf\tilde{Q}}}{\left(\frac{2}{3}{\rm Tr}{\bf\tilde{Q}}^2\right)^{3/2}}=\frac{\prod_{i=1}^3(\lambda_i-\bar{\lambda})}{\left(\frac{2}{3}\sum_{i=1}^{3}[\lambda_i-\bar{\lambda}]^2\right)^{3/2}}.
\end{equation}
  For a disk, $\Sigma=-1$, whereas for a  rigid rod $\Sigma=1$. Thus the  nature of asphericity has the bounds $-1 \le \Sigma \le 1$. Note that the so far described parameters $A_3$, $S$, and $\Sigma$ are all rotationally invariant and universal quantities \cite{aronovitz1986,cannon1991}.
  \begin{figure}[t!]
\centering
\includegraphics*[width=0.45\textwidth]{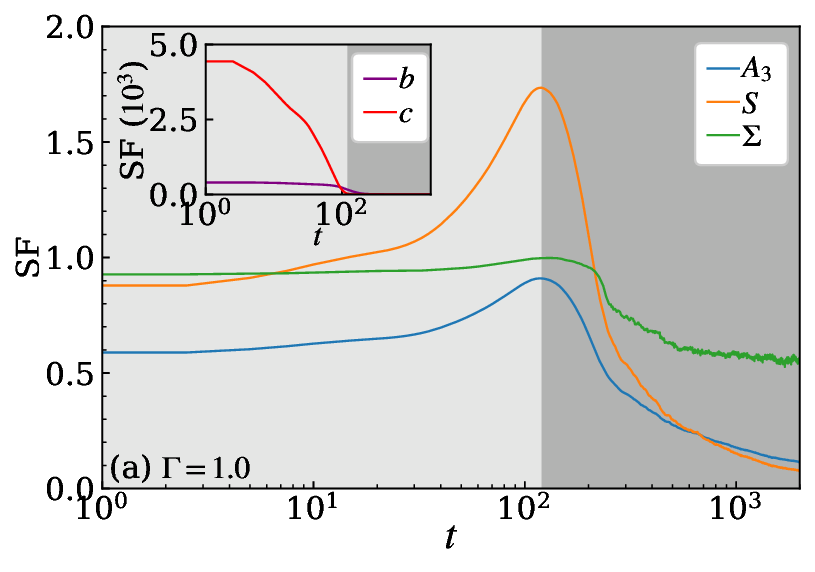}\\
\includegraphics*[width=0.45\textwidth]{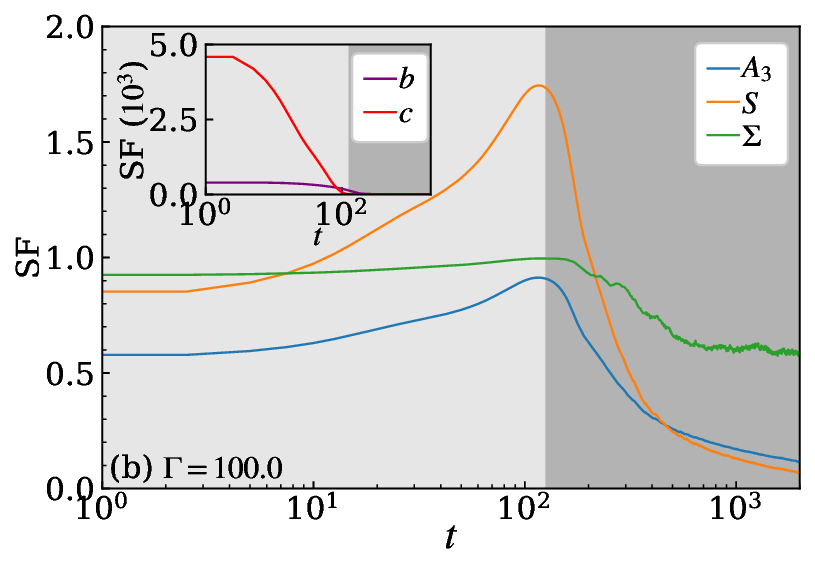}\\
\includegraphics*[width=0.45\textwidth]{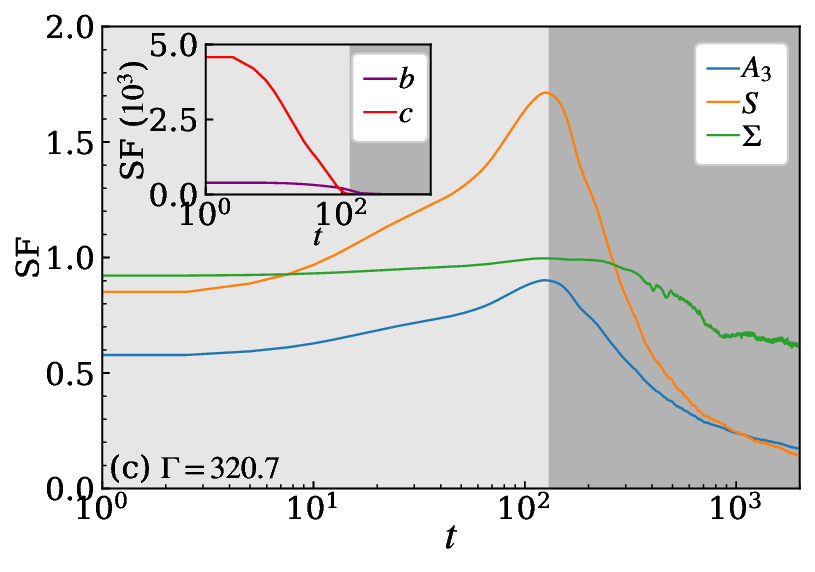}
\caption{\label{shape_test} Variation of shape factors (SF) $A_3$, $S$, and $\Sigma$ of a polymer of length $N=1024$ with time for three different values of the solvent viscosity $\Gamma$ at a temperature $T=0.3$. The inset shows the same variation for two more SFs, i.e., $b$ and $c$. For definitions of the SFs see the main text. The grey shades are used as guides to distinguish the pearl-necklace and sausage stages of the collapse. }
\end{figure} 
 
  \par
  Apart from the above parameters one can also consider certain parameters which are not invariant. For that one needs to sort the eigenvalues in descending order 
  $\lambda_1 > \lambda_2 > \lambda_3$ which allows us to define as a simple parameter the asphericity \cite{theodorou1985}
  \begin{equation}
   b=\frac{3}{2}\lambda_1 -\frac{1}{2}\sum_{i=1}^3\lambda_i=\lambda_1-\frac{1}{2}(\lambda_2+\lambda_3).
  \end{equation}
Similarly, as another simple parameter we calculate the acylindricity \cite{theodorou1985}
  \begin{equation}
 c=\lambda_2^2-\lambda_3^2.
\end{equation}
  For a perfectly spherical object, all eigenvalues are equal and hence $b = 0$, while a large positive value of $b$
indicates strong asphericity. For shapes of tetrahedral or even higher symmetry $b=c=0$, and for a perfect cylinder $c=0$.  Note that $b$ and $c$ can be
obtained only from the eigenvalues of ${\rm {\bf Q}}$.
\begin{figure}[t!]
\centering
\includegraphics*[width=0.5\textwidth]{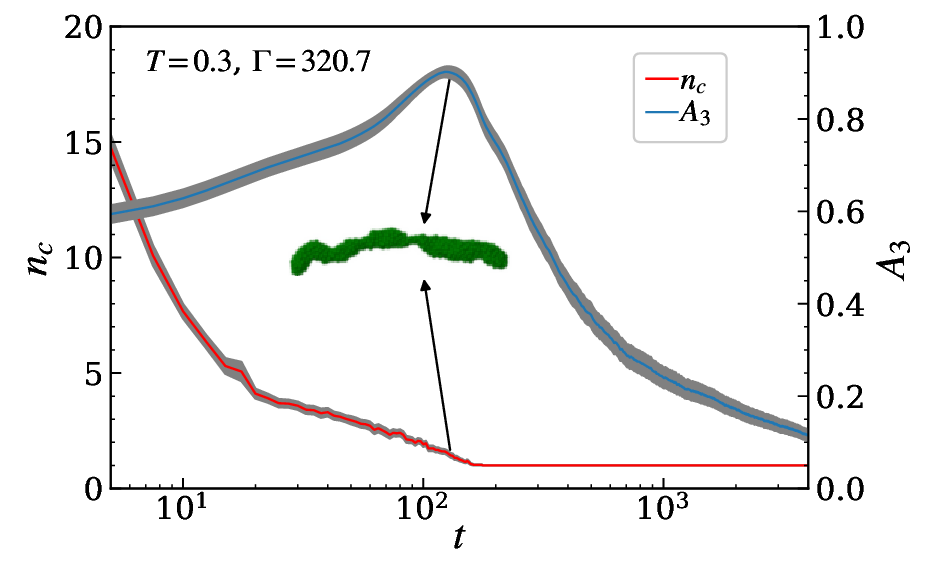}
\caption{\label{cluster_vs_A3} Time dependence of the number of clusters $n_c$ and the shape factor $A_3$ during the collapse of a polymer of length $N=1024$ at $T=0.3$ in a solvent of $\Gamma=320.7$. The snapshot represents a typical sausage conformation formed at the end of the pearl-necklace stage. The time where the $A_3$ data attain the peak value is a measure of the pearl-necklace relaxation time $\tau_p$. }
\end{figure}
\par
Utility of various shape factors in disentangling different stages of the collapse of a polymer in solvents of varying viscosity at $T=0.3$ is illustrated in Fig.\ \ref{shape_test}. In the main frame we show plots for the invariant shape parameters $A_3$, $S$, and $\Sigma$, while the simple but not invariant ones are shown in the insets. Both $A_3$ and $S$ appear to nicely capture the two distinct stages of the overall collapse process. They start off at the respective initial values for an extended coil. At very early time $t < 10$, the behaviors are almost flat marking the formation of stable clusters. In the range $t \in [10,150]$ both $A_3$ and $S$ show steady rise  highlighting the pearl-necklace stage. Eventually, they approach a maximum when all clusters have merged to form a single cluster, which at low $T$ resembles a sausage-like conformation. In the next stage the sausage starts relaxing via rearrangement of the monomers and then both $A_3$ and $S$ start decaying from their maximum values to approach zero, the
value for a perfectly spherical globule. Even though $\Sigma$ also shows a similar behavior as a function of time, it is difficult to pin-point the end of the pearl-necklace stage from it due to the absence of any prominent maxima in the data. Also at very late time $t > 10^3$ for all $\Gamma$, the data for $\Sigma$ almost saturate indicating no change in the conformation. However, by inspecting the conformation around that time it is clear that the conformation is still relaxing. This feature is clearly visible in the data of both $A_3$ and $S$. Thus, $A_3$ and $S$ will not only allow us to extract a time scale for the end of the pearl-necklace stage but can also potentially be used for estimating a relaxation time for the sausage stage. This will be done subsequently. The simpler quantities $b$ and $c$ also approach zero starting from a high value, reflecting the transition from the pearl-necklace to sausage stage. However, like $\Sigma$ both $b$ and $c$ do not show any significant features in the sausage stage that would allow one to extract a relaxation time for the sausage stage. Thus based on the above discussion it seems that both $A_3$ and $S$ are the two potential parameters that can be used to extract the relaxation times related to the different stages of the collapse. However, from now on we will be using only $A_3$ with the rational that $A_3 \approx 0$ in principle uniquely identifies a spherical object, whereas $S$ approaching its lower bound ideally indicates a disk-like object.

\par
To further substantiate our choice of $A_3$, we show in Fig.\ \ref{cluster_vs_A3} its compatibility with the time dependence of the number of clusters  $n_c$ of monomers formed in the pear-necklace stage of the collapse. For determining $n_c$, we first calculate the number of monomers present within a cut-off distance $r_c=2.5\sigma$ of the $i$-th monomer as
\begin{equation}
 n_i=\sum_{j=1}^N\Theta(r-r_{ij}),
\end{equation}
where $\Theta$ is the Heaviside step function. For, $n_i \ge n_{\rm min}$, we consider that there is a cluster
around the $i$-th monomer and all the $n_i$ monomers are part of that cluster. Here, we choose $n_{\rm min}=10$. This way the number of clusters may be overcounted, which we remove via a Venn diagram approach to find the actual number of discrete clusters $n_c$.  For the details on identifying and counting the number of clusters see Ref.\ \cite{majumder2020}. The data for $n_c$ are nicely synchronized  with the $A_3$ data presented in Fig.\ \ref{cluster_vs_A3} for a polymer of $N=1024$  at $T=0.3$ collapsing in a solvent of viscosity $\Gamma=320.7$. The point where the $A_3$ data attain its maximum nicely coincides with the point when $n_c=1$ for the first time. Thus, it is evident that $A_3$ can be relied on to extract the relaxation time marking the crossover from pearl-necklace to sausage-like phenomenology of the collapse. Thus we use the time where the $A_3$ data attain the peak value as a measure of the pearl-necklace relaxation time $\tau_p$. Following the pearl-necklace stage, the $n_c$ data remain constant at unity, but the $A_3$ data start decaying from its peak value and continue to do so characterizing the sausage relaxation.  
\begin{figure}[t!]
\centering
\includegraphics*[width=0.5\textwidth]{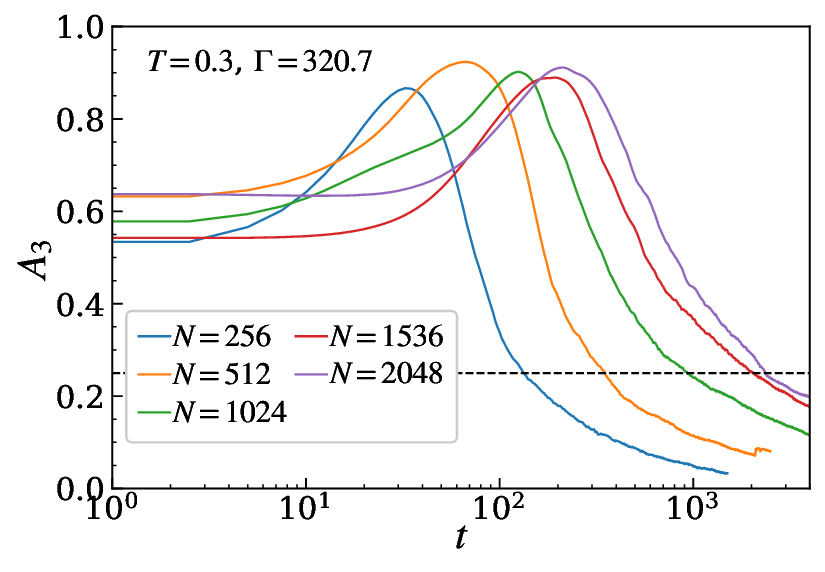}\\
\caption{\label{A3_diffN} Time dependence of $A_3$ for polymers of different lengths in a solvent of viscosity $\Gamma=320.7$ and temperature $T=0.3$. The dashed line represents the value $A_3=0.25$, used to determine the overall collapse time $\tau_c$. For an unambiguous visibility of the data, error bars are not shown here.}
\end{figure}
\par
In Fig.\ \ref{A3_diffN} we show variation of the time dependence of the $A_3$ data during the collapse with length of the polymer $N$ in a solvent with viscosity $\Gamma=320.7$ at $T=0.3$. One can clearly notice a monotonic dependence of the peak position with $N$, a signature of increasing $\tau_p$ with $N$ as expected. Similarly, one can also notice a slower decay of the $A_3$ data following the peak. This observation allows us to obtain an overall collapse time $\tau_c$ given as
\begin{equation}\label{tauc_measure}
 A_3(t=\tau_c)=h.
\end{equation}
We have checked that any value of $h$ in the range $0.1 \ge h\ge 0.45$ provides proportionate results for $\tau_c$. Subsequently, all the $\tau_c$ values presented are for the choice $h=0.25$, as represented by the dashed line in Fig.\ \ref{A3_diffN}. Now from the obtained $\tau_p$ and $\tau_c$ we determine the sausage relaxation time 
\begin{equation}\label{taus_measure}
 \tau_s=\tau_c-\tau_p.
\end{equation}

\begin{table}[b!]
\caption{Fitting results for the scaling of the pearl-necklace relaxation time $\tau_p$ with respect to the polymer length $N$ at different $T$ and $\Gamma$ using the ansatz in Eq.\ \eqref{taup_fit}.}

\begin{center}
\begin{tabular}{m{1.0cm} m{1.0cm} m{1.0cm} m{1.0cm}}
    \hline
    \hline
    \vskip 0.1cm
    $T$&$\Gamma$ &$\tau_p^0$ &~~~~~$z_p$ \\
    \hline
    \hline
    \vskip 0.1cm
    \multirow{3}{*}{\rotatebox[origin=c]{0}{$1.0$}}~~~~~~~~~~~~$1.0$ & ~~~~~~~~~~~~~$0.21(3)$ &~~~~~~~~~~~~~~~~~~~~~$0.97(3)$  \\  
  ~~~~~~~~~~~~~~$100.0$ & ~~~~~~~~~~~~~$0.22(3)$ &~~~~~~~~~~~~~~~~~~~~~$0.97(2)$\\ 
  ~~~~~~~~~~~~~~$320.7$ & ~~~~~~~~~~~~~$0.24(9)$ &~~~~~~~~~~~~~~~~~~~~~$0.96(1)$\\
   \hline
   \vskip 0.1cm
    \multirow{3}{*}{$0.3$}~~~~~~~~~~~~$1.0$ & ~~~~~~~~~~~~~$0.39(7)$ &~~~~~~~~~~~~~~~~~~~~~$0.83(2)$  \\  
  ~~~~~~~~~~~~~~$100.0$ & ~~~~~~~~~~~~~$0.23(3)$ &~~~~~~~~~~~~~~~~~~~~~$0.89(2)$ \\ 
  ~~~~~~~~~~~~~~$320.7$ & ~~~~~~~~~~~~~$0.23(5)$ &~~~~~~~~~~~~~~~~~~~~~$0.90(3)$ \\
    \hline
\end{tabular}
\end{center}
\label{fit_taup}
\end{table} 

\begin{table}[b!]
\caption{Fitting results for the scaling of the sausage relaxation time $\tau_s$ with respect to the polymer length $N$ at different $T$ and $\Gamma$ using the ansatz in Eq.\ \eqref{taus_fit}.}
\begin{center}
\begin{tabular}{m{1.0cm} m{1.0cm} m{1.0cm} m{1.0cm}}
    \hline
    \hline
    \vskip 0.1cm
    $T$&$\Gamma$ &$\tau_s^0$ &~~~~~$z_s$ \\
    \hline
    \hline
    \vskip 0.1cm
    \multirow{3}{*}{\rotatebox[origin=c]{0}{$1.0$}}~~~~~~~~~~~~$1.0$ & ~~~~~~~~~~~~~$0.61(9)$ &~~~~~~~~~~~~~~~~~~~~~$0.78(5)$  \\  
  ~~~~~~~~~~~~~~$100.0$ & ~~~~~~~~~~~~~$0.46(8)$ &~~~~~~~~~~~~~~~~~~~~~$0.83(3)$\\ 
  ~~~~~~~~~~~~~~$320.7$ & ~~~~~~~~~~~~~$0.90(9)$ &~~~~~~~~~~~~~~~~~~~~~$0.75(3)$\\
   \hline
   \vskip 0.1cm
    \multirow{3}{*}{$0.3$}~~~~~~~~~~~~$1.0$ & ~~~~~~~~~~~~~$0.06(3)$ &~~~~~~~~~~~~~~~~~~~~~$1.14(6)$  \\  
  ~~~~~~~~~~~~~~$100.0$ & ~~~~~~~~~~~~~$0.05(2)$ &~~~~~~~~~~~~~~~~~~~~~$1.19(6)$ \\ 
  ~~~~~~~~~~~~~~$320.7$ & ~~~~~~~~~~~~~$0.07(2)$ &~~~~~~~~~~~~~~~~~~~~~$1.20(4)$ \\
    \hline
\end{tabular}
\end{center}
\label{fit_taus}
\end{table} 
\begin{figure*}[t!]
\centering
\includegraphics*[width=0.875\textwidth]{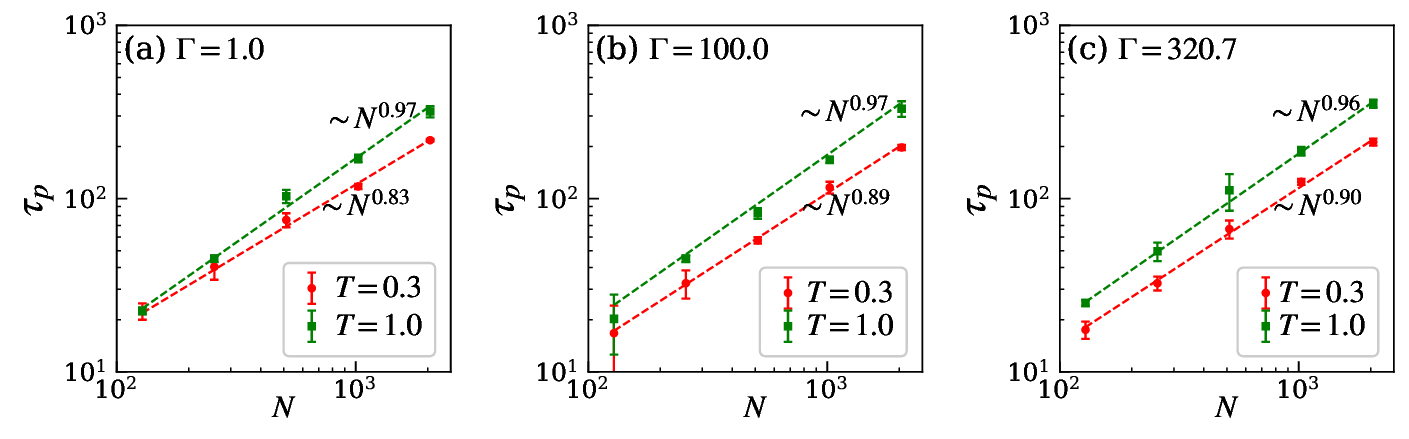}
\includegraphics*[width=0.875\textwidth]{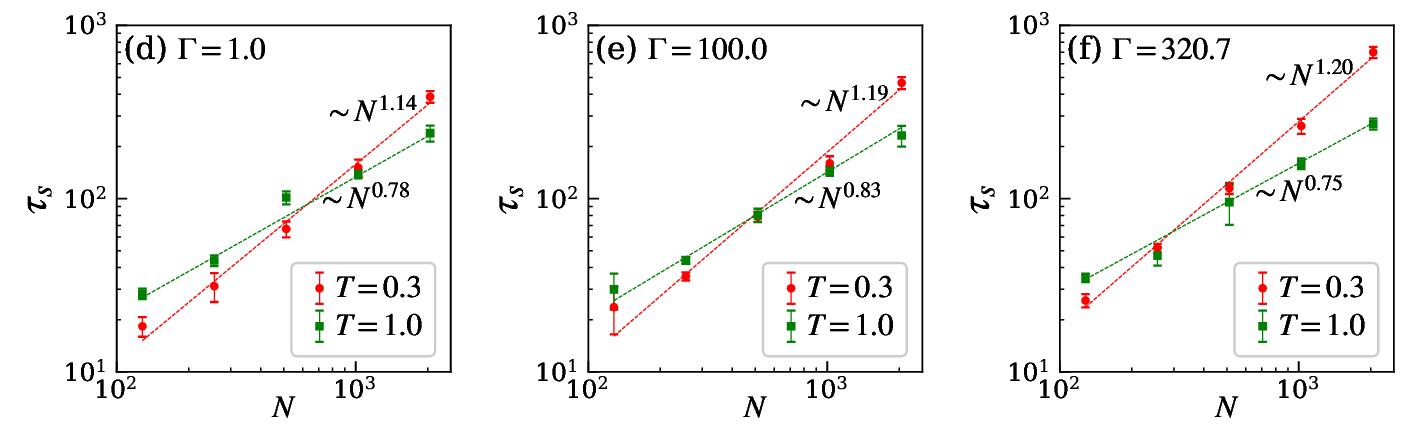}
\includegraphics*[width=0.875\textwidth]{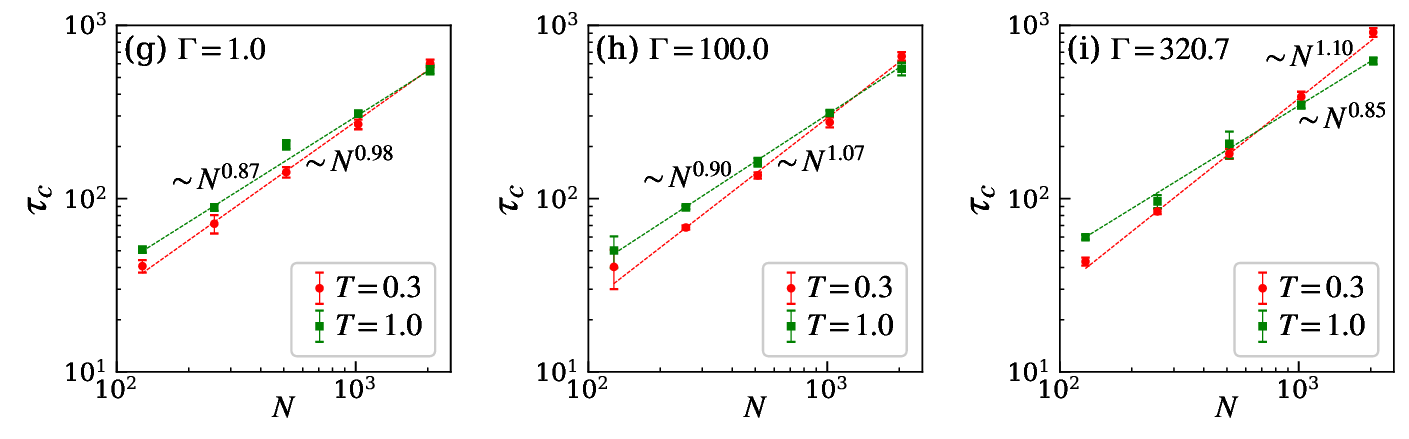}
\caption{\label{relaxation} Double-log plots showing the scaling of various relaxation times during the collapse as a function of the length of the polymer $N$ for different solvent viscosities. (a)-(c) are for the pearl-necklace relaxation time $\tau_p$, (d)-(f) are for the sausage relaxation time $\tau_s$, and (g)-(i) are for the overall collapse time $\tau_c$. Different dashed lines are best power-law fits whose exponents are indicated next to them. In each case, results for a low and a high $T$ are presented.}
\end{figure*}
\subsection{Scaling of relaxation times with chain length}\label{scaling_N}
With the methods for extracting various relaxation times in place, we first move on to investigate their scaling with respect to the length of the polymer. For that we use simulations at two fixed temperatures and three fixed solvent viscosities for five different chain lengths.  
\par
The obtained relaxation times  are presented in the log-log plots of Fig.\ \ref{relaxation} as a function of $N$, at a low and a high $T$ for different $\Gamma$. The data for the relaxation time $\tau_p$ characterizing the pearl-necklace stage, shown in the first row (a)-(c) show a power-law scaling of the form 
\begin{equation}\label{taup_fit}
 \tau_p=\tau_p^0N^{z_p},
\end{equation}
where $z_p$ is the corresponding dynamic exponent. The dashed lines in Figs.\ \ref{relaxation}(a)-(c) represent the best fit lines obtained after fitting the form in Eq.\ \eqref{taup_fit} to the data. The detail results of the fitting are presented in Table\ \ref{fit_taup}. It seems that $z_p$ is almost independent of $\Gamma$ at $T=1.0$. For $T=0.3$, the dynamics seems to be faster with smaller value of $z_p$, however, showing a slight increase as a function of $\Gamma$. 
The faster dynamics at low $T$ is an effect of the fact that the solvent becomes poorer with decreasing $T$, as previously also observed for the overall collapse time in cases where only the pearl-necklace stage is present in the collapse \cite{Majumder2017}. 
\begin{table}[b!]
\caption{Fitting results for the scaling of the overall collapse  time $\tau_c$ with respect to the polymer length $N$ using at different $T$ and $\Gamma$ using the ansatz in Eq.\ \eqref{tauc_fit}.} 
\begin{center}
\begin{tabular}{m{1.0cm} m{1.0cm} m{1.0cm} m{1.0cm}}
    \hline
    \hline
    \vskip 0.1cm
    $T$&$\Gamma$ &$\tau_c^0$ &~~~~~$z_c$ \\
    \hline
    \hline
    \vskip 0.1cm
    \multirow{3}{*}{\rotatebox[origin=c]{0}{$1.0$}}~~~~~~~~~~~~$1.0$ & ~~~~~~~~~~~~~$0.73(2)$ &~~~~~~~~~~~~~~~~~~~~~$0.87(5)$  \\  
  ~~~~~~~~~~~~~~$100.0$ & ~~~~~~~~~~~~~$0.61(3)$ &~~~~~~~~~~~~~~~~~~~~~$0.90(1)$\\ 
  ~~~~~~~~~~~~~~$320.7$ & ~~~~~~~~~~~~~$0.97(9)$ &~~~~~~~~~~~~~~~~~~~~~$0.85(1)$\\
   \hline
   \vskip 0.1cm
    \multirow{3}{*}{$0.3$}~~~~~~~~~~~~$1.0$ & ~~~~~~~~~~~~~$0.32(8)$ &~~~~~~~~~~~~~~~~~~~~~$0.98(6)$  \\  
  ~~~~~~~~~~~~~~$100.0$ & ~~~~~~~~~~~~~$0.18(4)$ &~~~~~~~~~~~~~~~~~~~~~$1.07(6)$ \\ 
  ~~~~~~~~~~~~~~$320.7$ & ~~~~~~~~~~~~~$0.19(3)$ &~~~~~~~~~~~~~~~~~~~~~$1.10(4)$ \\
    \hline
\end{tabular}
\end{center}
\label{fit_tauc}
\end{table} 
\par
The corresponding scaling plots of the sausage relaxation time $\tau_s$ are presented in Figs.\ \ref{relaxation}(d)-(f). There also the data seem to be following a power-law scaling which allows to perform a fitting using the ansatz
 \begin{equation}\label{taus_fit}
 \tau_s=\tau_s^0N^{z_s}.
\end{equation}
The resulting parameters are compiled in Table\  \ref{fit_taus}. For both $T$ the data are weakly dependent on the solvent viscosity with $z_s\in [0.75,0.83]$ and $z_s \in [1.14,1.20]$  for $T=1.0$ and $0.3$, respectively, indicating that the dynamics is faster at high $T$ unlike the pearl-necklace relaxation. This is due to fact that the relaxation in the sausage stage occurs via hydrodynamic dissipation of the surface energy, which gets accelerated with increasing  $T$ \cite{deGennes1985}.

\par
Finally we show the scaling of the overall collapse time $\tau_c$ in Figs.\ \ref{relaxation}(g)-(i). Again the data show a nice power-law scaling which we quantify using the usual ansatz of the form 
\begin{equation}\label{tauc_fit}
 \tau_c=\tau_c^0N^{z_c}.
 \end{equation}
 Results of the fitting exercise using \eqref{tauc_fit} are quoted in Table\ \ref{fit_tauc}. The data at $T=1.0$ indicate that the corresponding dynamic exponent $z_c \in [0.85,0.90]$ is nonmonotonic with the variation of $\Gamma$, whereas $z_c \in [0.98,1.10]$ at $T=0.3$ suggests a very weak dependence. 
 As a function of $\Gamma$ the trend in the data is similar to what is observed for $\tau_s$ suggesting that the sausage relaxation is the rate-limiting stage of the overall collapse process. Overall the data suggest that the scaling of $\tau_c$ also is dependent on $T$. The inclusion of solvent viscosity introduces a temperature-dependent frictional drag by the solvent molecules while the polymer tries to move, thus making the scaling of the relaxation times with the chain length a function of temperature. 
 \begin{figure*}[t!]
\centering
\includegraphics*[width=0.9\textwidth]{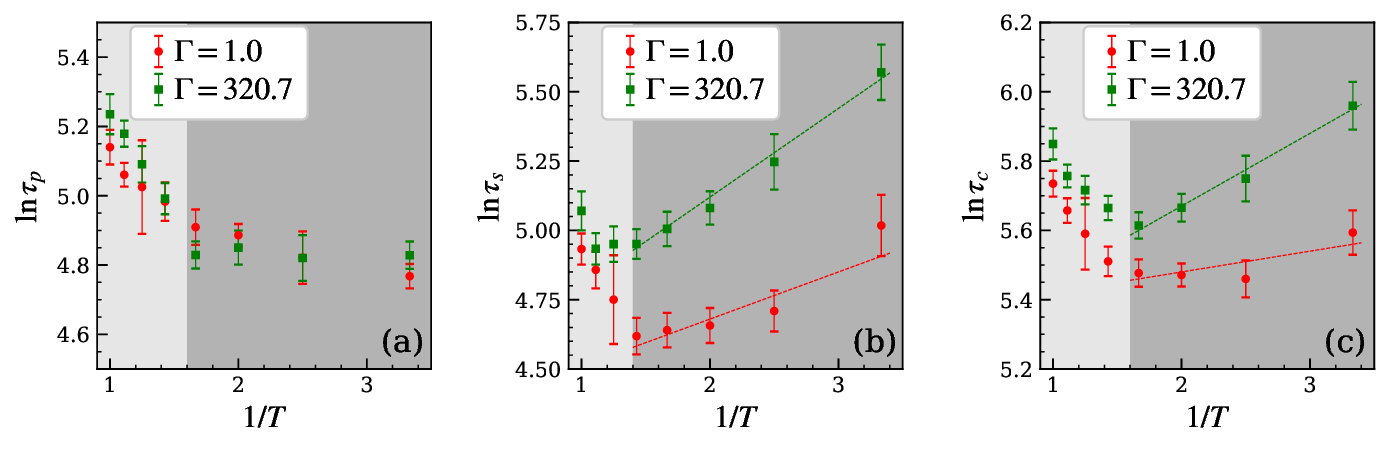}
\caption{\label{temp_dep} Temperature dependence of the relaxation times (a) $\tau_p$, (b) $\tau_s$, and (c) $\tau_c$ for a polymer of length $N=1024$ at two different solvent viscosities as indicated. We plot logarithm of the relaxation times as a function of the inverse temperature to check for  any Arrhenius behavior. In (a) the lighter shade marks the regime for anti-Arrhenius and the darker shade represents the temperature-independent regime. In (b) and (c), the lighter and darker shades, respectively, mark the regimes showing anti-Arrhenius and Arrhenius behaviors. The dashed lines in (b) and (c) show the best fit lines representing the respective Arrhenius behaviors according to Eq.\ \eqref{Arr_log}.}
\end{figure*}
\subsection{Temperature dependence of the relaxation times}
In this subsection we present results for the dependence of the relaxation times on temperature. For that we perform simulations at eight different $T$ using a polymer of fixed length $N=1024$ for two fixed solvent viscosities. 
\par
Considering the analogy between the collapse of a homopolymer and hydrophobic collapse 
or protein folding in general, it is to be noted that the folding rates $k$ of most proteins show Arrhenius behavior \cite{scalley1997}. Similarly, to classify structural glasses into fragile and non-fragile categories, one often checks whether the relaxation time $\tau$ follows an Arrhenius behavior. In terms of $\tau$ the Arrhenius behavior is given as 
\begin{equation}\label{Arr_lin}
 \tau=\tau_{\infty}\exp(E_a/k_BT),
\end{equation}
where $\tau_{\infty}$ is the amplitude and $E_a$ is the activation energy. On taking logarithm, Eq.\ \eqref{Arr_lin} transforms to 
\begin{equation}\label{Arr_log}
 \ln \tau=\ln \tau_{\infty}+\frac{E_a}{k_BT}.
\end{equation}
Thus, a linear behavior of the plot of $\ln \tau$ as a function of $1/T$ confirms an Arrhenius behavior. 
\begin{table}[b!]
\caption{Fitting results for the temperature dependence of the sausage relaxation time $\tau_s$ and the overall collapse time $\tau_c$ using the Arrhenius equation in Eq.\ \eqref{Arr_log}. } 
\begin{center}
\begin{tabular}{m{1.0cm} m{1.0cm} m{1.0cm} m{1.0cm}}
    \hline
    \hline
    \vskip 0.1cm
    $\tau$&$\Gamma$ &$\ln \tau_{\infty}$ &~~$E_a/k_B$ \\
    \hline
    \hline
    \vskip 0.1cm
    \multirow{2}{*}{\rotatebox[origin=c]{0}{$\tau_s$}}~~~~~~~~~~~~$1.0$ & ~~~~~~~~~~~~~~~~$4.34(9)$ &~~~~~~~~~~~~~~~~~~~~~$0.17(5)$  \\  
  ~~~~~~~~~~~$320.7$ & ~~~~~~~~~~~~~~~~$4.48(5)$ &~~~~~~~~~~~~~~~~~~~~~$0.32(2)$\\ 
     \hline
   \vskip 0.1cm
    \multirow{2}{*}{$\tau_c$}~~~~~~~~~~~~$1.0$ & ~~~~~~~~~~~~~~~~$5.36(8)$ &~~~~~~~~~~~~~~~~~~~~~$0.06(4)$  \\  
  ~~~~~~~~~~~$320.7$ & ~~~~~~~~~~~~~~~~$5.25(4)$ &~~~~~~~~~~~~~~~~~~~~~$0.21(2)$ \\ 
      \hline
\end{tabular}
\end{center}
\label{fit_temp}
\end{table} 

\par
In Fig.\ \ref{temp_dep} we present a test of the Arrhenius behavior of the relaxation times corresponding to the two stages of the collapse and the overall collapse time. 
 Over the considered $T$ range the data for $\tau_p$ in Fig.\ \ref{temp_dep}(a) show no signature of the Arrhenius behavior. At low $T$ it indicates no dependence on the temperature. On the other hand, in the high-$T$ range (marked by the lighter shade) it rather increases almost linearly with $T$ suggesting an anti-Arrhenius behavior, also observed in folding dynamics of certain proteins \cite{munoz1997,karplus2000,ferrara2000}. Since the range in $1/T$ is not sufficiently long we abstain ourselves from fitting  a straight line. Considering the fact that as $T$ increases the poor solvent approaches a good one, it is expected that the formation of small clusters or pearls of monomers slows down. 

\par
In the high-$T$ range similar behavior is observed for the sausage relaxation time $\tau_s$ presented in Fig.\ \ref{temp_dep}(b), however, the range of $T$ (marked by the lighter shade) seems to have shrinked. In the low-$T$ range the data for both $\Gamma$ clearly indicate a linear behavior with positive slope suggesting an Arrhenius behavior. This indulges us to fit Eq.\ \eqref{Arr_log} to the data, results of which are tabulated in Table\ \ref{fit_temp}. The dashed lines passing over the data points are the respective best fit straight lines obtained from this fitting exercise confirming the Arrhenius behavior. From the results obtained via fitting it appears that the amplitude $\tau_{\infty}$ is independent of $\Gamma$, whereas the activation energy for the sausage relaxation is larger for larger $\Gamma$.  
\par
 The data for the overall collapse time $\tau_c$ presented in Fig.\ \ref{temp_dep}(c) as a function of $T$ show almost the same behavior as the sausage relaxation time $\tau_s$. This again confirms that for the overall collapse the sausage relaxation is the rate-limiting stage. However, the high-$T$ range for the anti-Arrhenius behavior coincides with the data for $\tau_p$. Results from the fitting exercise with the Arrhenius behavior in Eq.\ \eqref{Arr_log} are also presented in Table\ \ref{fit_temp}. The corresponding best fit lines are represented by the dashed lines passing over the data points. The fit values obtained once again infer that the amplitude is independent of the solvent viscosity $\Gamma$. The corresponding activation energy $E_a$  is apparently larger for larger $\Gamma$. According to the error bars, the range of the ratio of $E_a$ obtained for $\Gamma=1.0$ and $\Gamma=320.7$ varies from $1.36$ to $2.83$ for the values extracted from $\tau_s$. For the values extracted from $\tau_c$, the corresponding ratio spans a much wider range from $1.90$ to $11.5$. Due to the wider range for $\tau_c$ it would not be advisable to extract the activation energy from it. On the other hand, the $E_a$ values extracted from $\tau_s$ with a narrow range provides a more reliable measure of the activation energy of the overall collapse process. This can also be appreciated from the fact that the sausage relaxation is the bottle neck of the overall collapse process. 
 \begin{figure*}[t!]
\centering
\includegraphics*[width=0.9\textwidth]{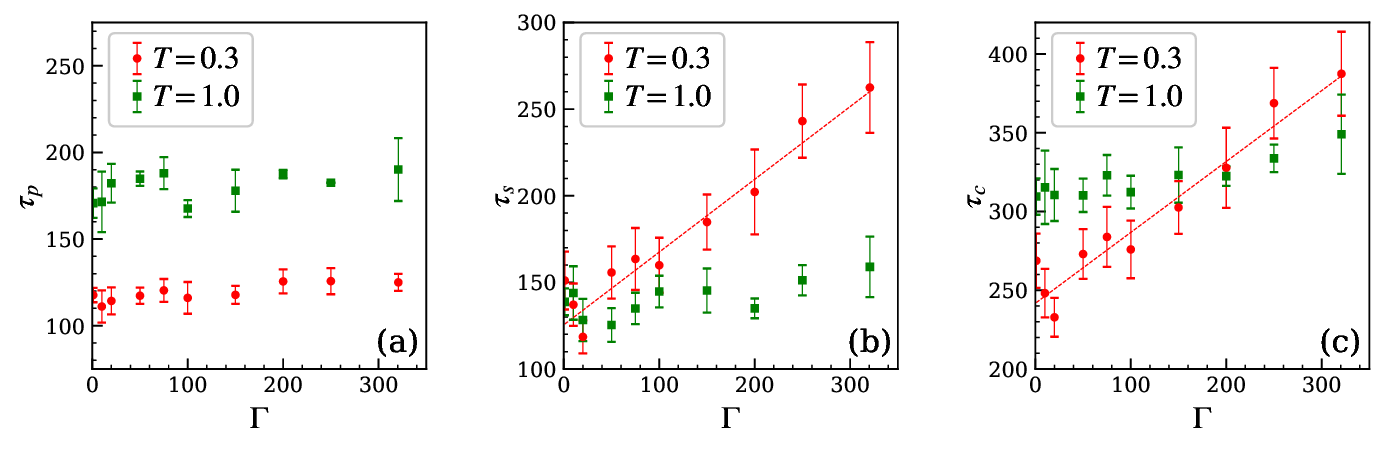}
\caption{\label{tau_Gamma} Solvent viscosity dependence of the relaxation times (a) $\tau_p$, (b) $\tau_s$, and (c) $\tau_c$ for a polymer of length $N=1024$ at two different temperatures as indicated. The dashed line passing over the data for $T=0.3$ in (b) is the best fit using Eq.\ \eqref{taus_intrinsic} with $\tau_s^{\rm int}=125.5$ and $\tau_s^{\rm sol}=0.42$. Similarly, the dashed line in (c) is the best fit using Eq.\ \eqref{tauc_intrinsic} with $\tau_c^{\rm int}=241.73$ and $\tau_c^{\rm sol}=0.45$.}
\end{figure*}
 \subsection{Solvent-viscosity dependence of the relaxation times}
 Finally we present results for the dependence of the relaxation times on the solvent viscosity $\Gamma$. For that we perform simulations at ten different $\Gamma$ for a polymer of fixed length $N=1024$ at two fixed temperatures. 
 \par
 The obtained dependence of the relaxation times as a function of $\Gamma$ are shown in Fig.\ \ref{tau_Gamma}. The pearl necklace relaxation time $\tau_p$ presented in Fig.\ \ref{tau_Gamma}(a) shows no dependency on the solvent viscosity $\Gamma$ for both temperatures. This can be intuitively derived from the fact that in the pearl-necklace stage local ordering of the monomers is the driving force which presumably  depends negligibly on the solvent viscosity. On the other hand, the sausage relaxation occurs via the minimization of the surface energy which gets opposed by the motion of the solvent molecules, unless there are sufficient thermal fluctuations to cross that barrier. This is reflected by the dependence of the sausage relaxation time $\tau_s$ on the solvent viscosity $\Gamma$, presented  in Fig.\ \ref{tau_Gamma}(b). The apparent linear behavior of the data for $T=0.3$ calls for a fit using the following ansatz  
 \begin{equation}\label{taus_intrinsic}
  \tau_s=\tau_s^{\rm int}+\tau_s^{\rm sol}\Gamma,
 \end{equation}
 where $\tau_s^{\rm int}$ represent the relaxation time due to the internal friction of the polymer and $\tau_s^{\rm sol}$ is the amplitude for the dependence on the solvent viscosity $\Gamma$. The ansatz in Eq.\ \eqref{taus_intrinsic} is commonly used for viscosity dependence of folding rates of proteins \cite{qiu2004,schulz2012,zheng2015}. There, the linear behavior is originally motivated by drawing analogy between protein folding and chemical reactions in solution, and thereby using the Kramer's theory of linear dependency of folding relaxation times and the solvent viscosity \cite{kramers1940}. The dashed line in Fig.\ \ref{tau_Gamma}(b) is the best fit resulting from fitting Eq.\ \eqref{taus_intrinsic} with the data for $T=0.3$. The corresponding values are $\tau_s^{\rm int}=125.5(5.9)$ and $ \tau_s^{\rm sol}=0.42(5)$. One of course could argue about a fit using a pure power-law of the form $\tau_s \sim \Gamma^{\kappa}$. For protein folding it has been argued \cite{qiu2004,schulz2012} that such a power-law dependence requires the relaxation rate to diverge at vanishing solvent viscosity, whereas an ansatz of the form \eqref{taus_intrinsic} implies that another damping mechanism intercedes to control the folding rate as friction due to solvent vanishes. With hindsight and topped up  by the apparent linear behavior of $\tau_s$ with $\Gamma$ at $T=0.3$, we abstain ourselves from using a power-law fit to the data. At $T=1.0$ the data for $\tau_s$ shows negligible dependence on $\Gamma$ since the thermal fluctuations are sufficiently strong to overcome the barrier due to friction of solvent molecules.
\par
In Fig.\ \ref{tau_Gamma}(c) we show the dependence of the overall collapse time $\tau_c$ on the solvent viscosity $\Gamma$. Since the sausage relaxation is the rate-limiting stage of the overall collapse process, the behavior of $\tau_c$ is similar to what is observed for $\tau_s$. This compels us to analyze the data at $T=0.3$ using the linear ansatz analogous to \eqref{taus_intrinsic}
\begin{equation}\label{tauc_intrinsic}
  \tau_c=\tau_c^{\rm int}+\tau_c^{\rm sol}\Gamma.
 \end{equation}
A fit with the above form to the data yields 
$\tau_c^{\rm int}=241.73(6.3)$ and $\tau_c^{\rm sol}=0.45(5)$. Note that the obtained $\tau_c^{\rm sol} \approx \tau_s^{\rm sol}$ suggesting that the amplitude for dependence on solvent viscosity is rather robust. The data for $T=1.0$ show negligible dependence on $\Gamma$ as observed for $\tau_s$.
\section{Conclusion}\label{conclusion}
We have presented results exploring the nonequilibrium pathways of the collapse transition of a flexible homopolymer when it is quenched from its extended coil state in a good solvent to a poor solvent where the equilibrium conformation is a compact globule. We have performed simulations considering explicit solvent molecules in a framework using the Lowe-Andersen thermostat which allowed us to tune the solvent viscosity $\Gamma$. Results from our simulations over a wide range of $\Gamma$ and at different temperatures $T$ revealed new insights about the collapse pathway. For moderate to high $T$ the collapse occurs through pearl-necklace intermediates in accordance with the Halperin-Goldbart picture, independent of the solvent viscosity as observed in previous simulations without considering solvent viscosity. Interestingly, at low $T$ our results confirm the de Gennes prediction that following the pearl-necklace stage the polymer attains a distinct sausage-like conformation which subsequently, via minimization of the surface energy, approaches a compact spherical globule. Presence of the sausage-like intermediates gets prolonged with increase of the solvent viscosity $\Gamma$. This brings us to the conjecture that the collapse pathway of a polymer is a function of the temperature and solvent viscosity, depending on which one observes initially the pearl-necklace intermediate followed by the  sausage intermediates. 
\par
Since the usual protocol of understanding the collapse dynamics was not adequate, we have used different shape factors derived from the gyration tensor to disentangle the pearl-necklace and the sausage relaxation stages. This also allowed us to extract the respective relaxation times of the two stages of the collapse as well as the overall collapse time. Both the pearl-necklace relaxation time $\tau_p$ and the sausage relaxation time $\tau_s$ show power-law scaling as a function of the polymer length $N$. A smaller value of the power-law exponent $z_p$ for $\tau_p$ at low $T$ suggests a faster dynamics of the pearl-necklace stage as a courtesy of the fact that the solvent becomes poorer with decrease in $T$. Conversely, the power-law exponent $z_s$ indicates that the dynamics  of the sausage relaxation is faster at higher $T$. This is in accordance with de Gennes phenomenological picture suggesting that  the minimization of the surface energy of the sausage-like intermediates occurring via  hydrodynamic dissipation of energy is faster at high $T$. The overall collapse time $\tau_c$ shows similar scaling behavior as $\tau_s$ implying that the sausage relaxation is the rate-limiting stage in the overall collapse process of the polymer. 
\par
We have shown that for a fixed $N$, all the relaxation times show an anti-Arrhenius behavior in  the high-$T$ range. In the low-$T$ regime, whereas $\tau_p$ is almost insensitive to $T$, both $\tau_s$ and $\tau_c$ follow Arrhenius behavior from where one can extract the corresponding activation energy of the sausage relaxation and the overall collapse. For a fixed $T$ and $N$, results for varying solvent viscosity $\Gamma$ showed that the pearl-necklace relaxation dynamics is independent of $\Gamma$. At low $T$, dependence of $\tau_s$ on $\Gamma$ have shown a linear behavior, similar to what is found for folding rates of proteins. The agreement of the data to a straight line with an off-set is further indicative of the existence of internal friction of the polymer. This implies that internal friction of proteins could plausibly originating from its polymeric backbone rather than depending on the specifics of its sequence. A study of semiflexible heteropolymers using the framework of the present work would be able to shed more light on this, which we take up as a our next endeavor \footnote{H. Christiansen, S. Majumder, and W. Janke, work in progress}. 
\par
In future, it would be interesting to study the dependence of the cluster coarsening on the solvent viscosity. Intriguing would also be to investigate the role of solvent viscosity in the aging and related scaling during the collapse. Polymers are found to be residing in a different, confined geometrical setup in the real world. Hence, it would be worth exploring both the equilibrium and nonequilibrium dynamics of a polymer in restricted geometries using our current simulation setup. Going by the recent trend of investigating biomolecules as active entities, it would certainly be promising to study active polymers in this explicit solvent setup where the solvent viscosity can be tuned. 
\acknowledgments
S.M.\ was funded by the Science and Engineering Research Board (SERB), Govt.\ of India through a Ramanujan Fellowship (File no.\ RJF/2021/000044). W.J.\ thanks funding by the Deutsche Forschungsgemeinschaft (DFG, German Research Foundation) – 469\ 830\ 597 under project ID JA\ 483/35-1.

%Control: production of eprint (-1) disabled
%

% \bibliography{bib_new.bib}
\end{document}